%% file: beastly.tex
\documentclass{sigchi}

\CopyrightYear{2019}
\setcopyright{acmlicensed}
\doi{https://doi.org/10.1145/3311350.3347172}
\isbn{978-1-4503-6688-5/19/10}
\conferenceinfo{CHI PLAY'19,}{October  22--25, 2019, Barcelona, Spain}
\acmPrice{\$15.00}

\usepackage{balance}       
\usepackage{graphics}      
\usepackage[T1]{fontenc}   
\usepackage{txfonts}
\usepackage{mathptmx}
\usepackage[pdflang={en-US},pdftex]{hyperref}
\usepackage{color}
\usepackage{booktabs}
\usepackage{textcomp}

\usepackage{microtype}        
\usepackage{ccicons}          

\usepackage{todonotes}

\def\plaintitle{Beyond Human: Animals as an Escape from \\ Stereotype Avatars in Virtual Reality Games}

\def\emptyauthor{}
\def\plainkeywords{Animal avatars; virtual creatures; animal embodiment; IVBO; virtual reality games; avatar control.}

\makeatletter
\def\url@leostyle{%
  \@ifundefined{selectfont}{
    \def\UrlFont{\sf}
  }{
    \def\UrlFont{\small\bf\ttfamily}
  }}
\makeatother
\def\pprw{8.5in}
\def\pprh{11in}

\definecolor{linkColor}{RGB}{6,125,233}
\hypersetup{%
  pdftitle={\plaintitle},
  pdfauthor={\emptyauthor},
  pdfkeywords={\plainkeywords},
  pdfdisplaydoctitle=true, 
  bookmarksnumbered,
  pdfstartview={FitH},
  colorlinks,
  citecolor=black,
  filecolor=black,
  linkcolor=black,
  urlcolor=linkColor,
  breaklinks=true,
  hypertexnames=false
}
\usepackage{tabularx}
\usepackage{threeparttable}
\usepackage{enumitem}
\include{ivda-macros}

\begin{document}
\teaser{
  \includegraphics[width=\linewidth]{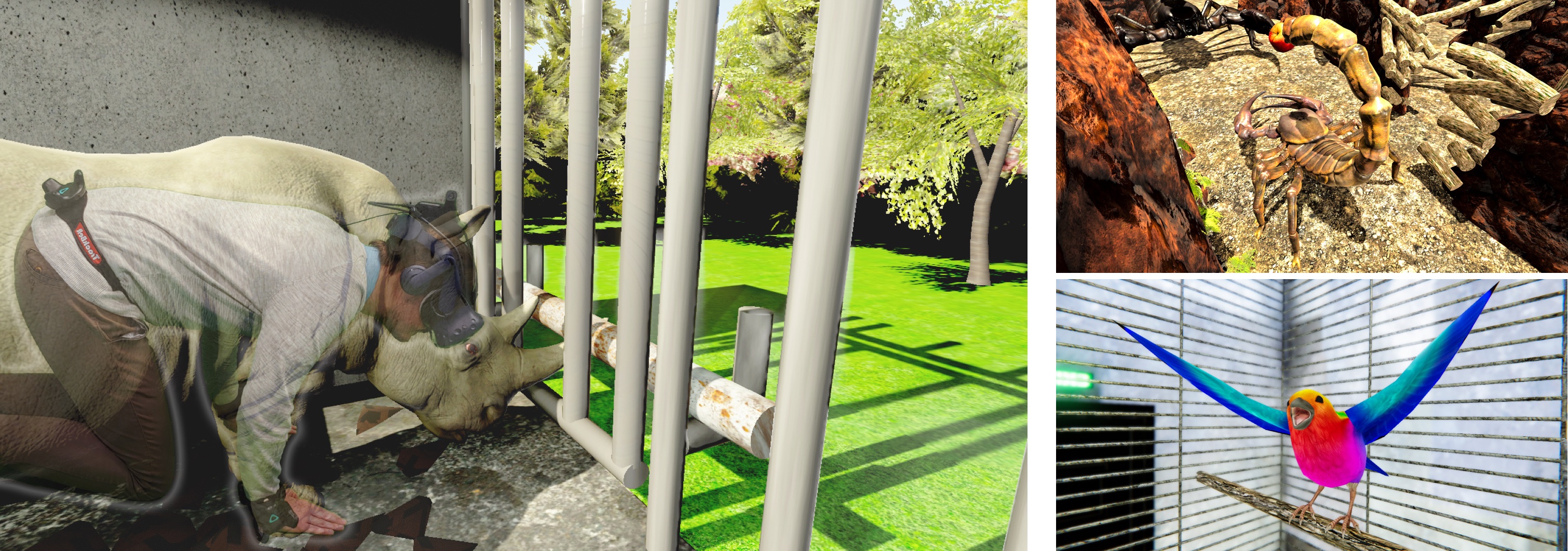}
  
  \caption{We explore the potential of nonhuman avatars in VR games. The evaluation of our three escape room games for different animal types reveals that players enjoy the control over additional body parts, as such morphologies allow novel, refreshing interactions and enable superhuman abilities.}
  \label{fig:teaser2}}

\title{\plaintitle}

  \numberofauthors{1}
\author{
  \alignauthor Andrey Krekhov$^{1}$, Sebastian Cmentowski$^{1}$, Katharina Emmerich$^{2}$, Jens Kr\"uger$^{1}$ \\
    \affaddr{$^{1}$High Performance Computing Group, $^{2}$Entertainment Computing Group}\\
    \affaddr{University of Duisburg-Essen, Germany}\\
    \email{\{andrey.krekhov, sebastian.cmentowski, katharina.emmerich, jens.krueger\}@uni-due.de}\\
  }

\maketitle

\begin{abstract}

Virtual reality setups are particularly suited to create a tight bond between players and their avatars up to a degree where we start perceiving the virtual representation as our own body. We hypothesize that such an illusion of virtual body ownership (IVBO) has a particularly high, yet overlooked potential for nonhumanoid avatars. To validate our claim, we use the example of three very different creatures---a scorpion, a rhino, and a bird---to explore possible avatar controls and game mechanics based on specific animal abilities. A quantitative evaluation underpins the high game enjoyment arising from embodying such nonhuman morphologies, including additional body parts and obtaining respective superhuman skills, which allows us to derive a set of novel design implications. Furthermore, the experiment reveals a correlation between IVBO and game enjoyment, which is a further indication that nonhumanoid creatures offer a meaningful design space for VR games worth further investigation.

\end{abstract}



\section{Introduction}

The choice of our virtual representation, our avatar, has a strong influence on how we perceive a game. Hence, introducing novel avatar kinds, beyond stereotypes such as knights and wizards, is a viable option to create refreshing and engaging player experiences. This choice applies even more for virtual reality (VR) games, because such immersive setups are capable of amplifying the bond with our virtual self. That bond can be strong enough such that we start perceiving the virtual representation as our own body---a phenomenon also known as the \textit{illusion of virtual body ownership (IVBO)}~\cite{slater2010first}. 

By a smart choice of avatars, VR games could allow us to collect impressions and experiences that would not be possible or would be far less engaging in a nonimmersive setup. One prominent example is games focused on nonhumanoid creatures, be it real animals or mythical creatures. Even though players enjoy ``beastly'' non-VR games, such as \textit{Black \& White}~\cite{BlackWhite} and \textit{Deadly Creatures}~\cite{Deadly}, similar scenarios are offered very rarely. Especially in VR, where presence and the IVBO effect could significantly intensify our experience when using animal abilities, games like \textit{Eagle Flight}~\cite{Eagle} remain an exception.

We see manifold reasons why that potential remains unfulfilled, including the very few studies on creature embodiment in VR, which makes it difficult for game designers to predict whether and how players will perceive animal avatars. Furthermore, as only a few games have touched upon this topic, best practices and design guidelines for such avatars are lacking. In other words, we need further research to understand the challenges and opportunities induced by the nonhuman morphology, e.g., additional limbs and their influence on IVBO, differing postures, and possible control approaches.

Our paper makes two contributions. First, we explore nonhumanoid avatars in VR using escape room games built around three very different animals: a rhino, a scorpion, and a bird (cf. \FG{fig:teaser2}). Each game explores a different control mechanism and focuses on distinct ``superhuman'' skills that are typical for these animals. Our evaluation underpins the resulting high player enjoyment, especially from these animal abilities and additional body parts, such as horns, tails, or wings. Accordingly, we draw design implications for animal avatars and present our lessons learned during the design of such VR games.

Our second contribution is the investigation of IVBO in such scenarios. We study how the nonhuman morphology influences our ability to embody such avatars in VR games. In particular, our evaluation reveals correlations between IVBO, game enjoyment, and presence, and confirms that additional body parts and skills are not an obstacle for inducing IVBO. Hence, we assume that our work will motivate researchers and practitioners to reconsider IVBO-enabled nonhumanoid avatars as an important component of player experience in VR.


\section{Related Work}

As our research targets virtual environments, we begin with a brief introduction of the related VR terms before focusing on the embodiment of nonhumanoid avatars. Nowadays, VR has regained attention mostly because of affordable mainstream HMDs, such as HTC Vive~\shortcite{vive}, which allow players to experience games from a novel perspective. Thereby, researchers~\cite{Biocca:1995:IVR:207922.207926,sherman2002understanding} usually refer to \textit{immersion}~\cite{cairns2014immersion} as the technical quality of VR equipment and apply the term \textit{presence}~\cite{slater1995taking, slater2003note} to describe the impact of such devices on our perception. In our case, we are particularly interested in presence, which can be measured as proposed by, e.g., IJsselsteijn et al.~\shortcite{IJsselsteijn} and Lombard and Ditton~\shortcite{lombard1997heart}.

Immersive technologies not only allow us to experience such a ``feeling of being there''~\cite{heeter1992being}, but also increase our ability to emphasize our virtual self-representation. We can embody our avatar to a remarkable degree, which is also referred to as the \textit{illusion of virtual body ownership (IVBO)}~\cite{lugrin2015anthropomorphism}, agency, or body transfer illusion.

IVBO originates in the effect of body ownership. The initial experiments by Botvinick and Cohen~\shortcite{botvinick1998rubber} introduced the rubber hand illusion: the participant's arm was hidden and replaced by an artificial rubber limb, and stroking both the real and virtual arms created the illusion of actually owning that artificial limb. After further investigations~\cite{tsakiris2005rubber}, researchers proposed a number of models~\cite{tsakiris2010my,ehrsson2007experimental,petkova2008if,lenggenhager2007video} to explain such an interplay between external stimuli and our internal body perception.

Slater et al.~\shortcite{slater2008towards} and Banakou et al.~\shortcite{banakou2013illusory} transferred the original body ownership effect, including the underlying visuotactile stimulation, to virtual environments. However, in their later work, Slater et al.~\cite{slater2010first} and Sanchez-Vives et al.~\cite{sanchez2010virtual} revisited the stimuli correlations and concluded that sensorimotor cues are more important than the visuotactile cues, which is an important insight, as VR setups seldom include tactile stimulations. To complete the picture, apart from visuotactile and sensorimotor cues, the IVBO effect is mainly impacted by visuoproprioceptive cues (perspective, body continuity, posture and alignment, appearance, and realism)~\cite{slater2009inducing,slater2010first,perez2012my,maselli2013building}.

IVBO was mainly explored with anthropomorphic characters and realistic representations~\cite{lugrin2015anthropomorphism, lin2016need,jo2017impact}. For instance, related to the question of avatar customization in games, Waltemate et al.~\shortcite{waltemate2018impact} showed that customizable representations lead to significantly higher IVBO effects. 

A strong IVBO can produce various changes in player behavior~\shortcite{jun2018full,muller2017through}, resembling the Proteus Effect by Yee et al.~\shortcite{yee2007proteus}. For instance, the work by Peck et al.~\shortcite{peck2013putting} revealed a significant reduction in racial bias when players embody a black character. Similarly, virtual race can also affect the drumming style~\cite{kilteni2013drumming}. Other reactions are childish feelings arising from embodying child bodies~\cite{banakou2013illusory} and an increase in perceived stability when having a robotic avatar~\cite{lugrin2016avatar}. Hence, prior work indicates that IVBO can be used to evoke specific feelings and attributes~\cite{kors2016breathtaking}. We suggest that a strong bond to the creature caused by IVBO can also increase our involvement with environmental issues~\cite{ahn2016experiencing,berenguer2007effect} and our empathy for animals, which, in turn, is transferable to human-human empathy, as shown by Taylor et al.~\shortcite{taylor2005empathy}.

Researchers have also expressed interest in studying IVBO beyond human morphology. For instance, Riva et al.~\shortcite{riva2014interacting} posed the following question: \textit{But what if, instead of simply extending our morphology, a person could become something else- a bat perhaps or an animal so far removed from the human that it does not even have the same kind of skeleton— an invertebrate, like a lobster?} Interestingly, embodying a bat is even being discussed in philosophy~\cite{nagel1974like}. If we consider exotic body compositions, as in the case of a lobster, that have few properties in common with our human body, the idea of sensory substitution~\cite{bach2003sensory} might play an important role. One might also consider such substitution mechanisms as playful interactions: e.g., the echolocation feature of a bat could be replaced by tactile feedback in a VR game.

Given the extreme diversity of real and fictional creatures, it is difficult or even impossible to research IVBO for virtual animals as a whole. Instead, previous research tackled isolated modifications of body parts. For instance, Kilteni et al.~\shortcite{kilteni2012extending} were able to stretch the virtual arm up to four times its original length without losing IVBO. Normand et al.~\shortcite{normand2011multisensory} used IVBO to induce the feeling of owning a larger belly than in reality. As a first step toward generalization, Blom et al.~\shortcite{blom2014effects} concluded that strong spatial coincidence of real and virtual body part is not mandatory to produce IVBO.

Certain animals, such as scorpions or rhinos in our study, have additional body parts that players might want to control. In this respect, prior work~\cite{ehrsson2009many,guterstam2011illusion} confirmed that having an additional arm preserves IVBO and induces a double-touch feeling. Steptoe et al.~\shortcite{steptoe2013human} reported effects of IVBO upon attaching a virtual tail-like body extension to the user’s virtual character. Clearly, these findings are relevant for a plethora of real and fictive nonhumanoids, such as dragons. The authors also discovered higher degrees of IVBO when the tail movement is synchronized with the real body.

To remain briefly with the example of a dragon as an avatar: Egeberg et al.~\shortcite{Egeberg:2016:EHB:2927929.2927940} proposed different ways wing control could be coupled with sensory feedback, and Sikstr\"om et al.~\shortcite{sikstrom2014role} assessed the influence of sound on IVBO in such scenarios.  Won et al.~\shortcite{won2015homuncular} further analyzed our ability to inhabit nonhumanoid avatars that have additional body parts.

Closely related to our research, the works-in-progress paper by Krekhov et al.~\shortcite{krekhov2018anim} also suggested embodying virtual animals in VR games. In their preliminary, explorative study, the authors implemented different control approaches for virtual tigers, bats, and spiders, and reported tendencies that IVBO remains intact for such avatars. We continue that work by building on the lessons learned regarding full-body and half-body control approaches, yet focus on embedding this knowledge into games research.

Naturally, we need a way to measure and compare IVBO strength in order to investigate whether and how IVBO influences player experience. In this regard, we point readers to the recent work by Roth et al.~\shortcite{roth2017alpha} that introduced the \textit{alpha IVBO questionnaire} based on a fake mirror scenario study. The authors suggested acceptance, control, and change as the three factors that determine IVBO. As the study by Krekhov et al.~\shortcite{krekhov2018anim} relied on this questionnaire to study animal embodiment in VR, we  applied the same process to generate comparable results.

A body of literature related to the control of animal avatars should be mentioned in this context. Leite et al.~\shortcite{leite2012shape} experimented with virtual silhouettes of animals that were used like shadow puppets and controlled by body motion. For 3D cases, Rhodin et al.~\shortcite{rhodin2014interactive} applied sparse correspondence methods to create a mapping between player movements and animal behavior and tested their approach with species such as spiders and horses. As a next step, Rhodin at al.~\shortcite{rhodin2015generalizing} experimented with the generalization of wave gestures to create control possibilities for, e.g., caterpillar crawling movements. Our research extends these methods by presenting additional mechanisms tailored to animal avatar control.

\begin{figure}[t!]
\centering
\includegraphics[width=1.0\columnwidth]{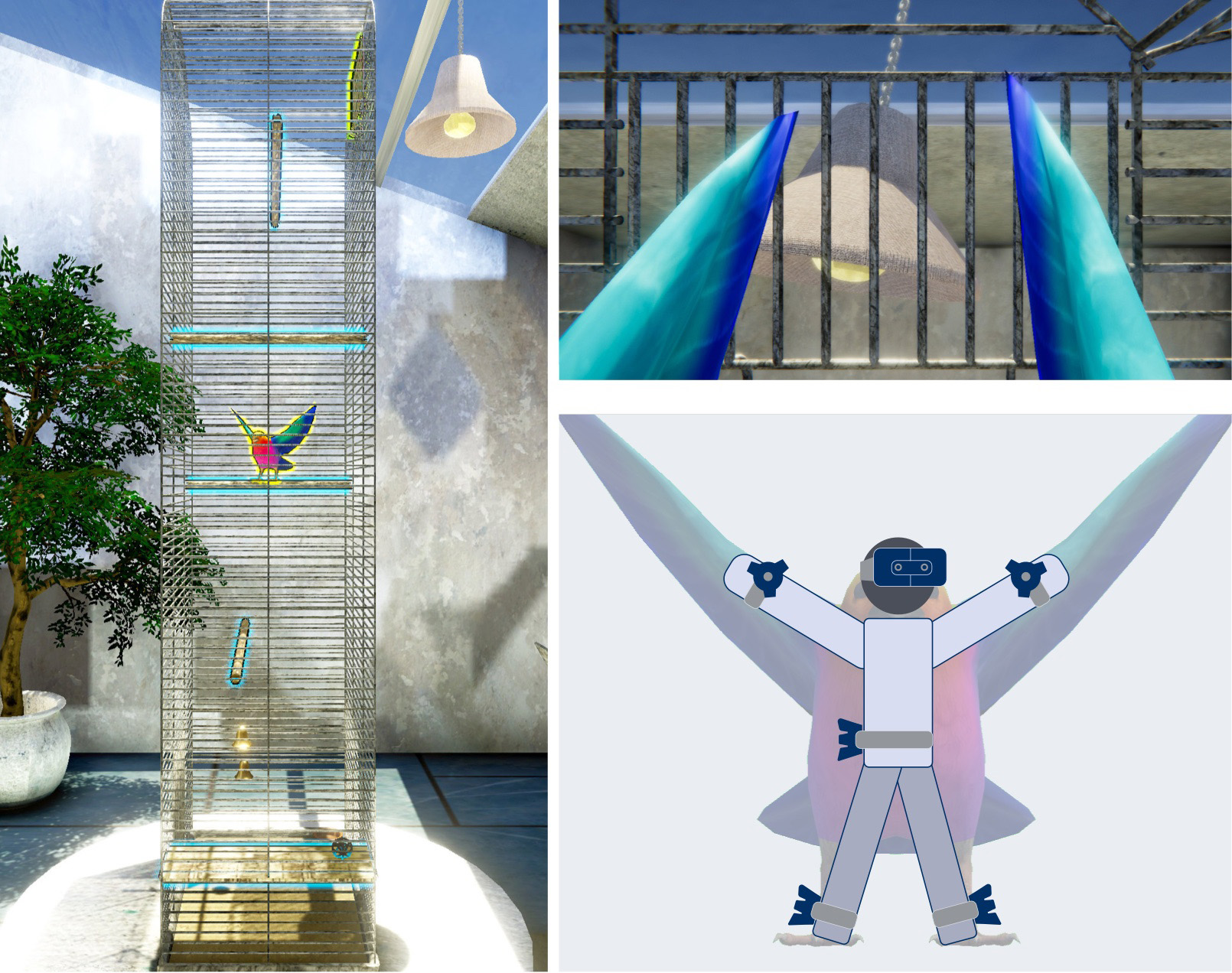}
\caption{
\textit{Bird Cage.} Players embodied a bird that was caught in a cage and had to escape through the top right door (marked yellow). The blue marked rods could be used for rests between the exhausting flights performed by full-body controls (bottom right). Finally, wind gusts had to be created to turn a lamp into a wrecking ball (top right).}
\label{fig:bird}
\end{figure}

\begin{figure*}[t!]
\centering
\includegraphics[width=2.1\columnwidth]{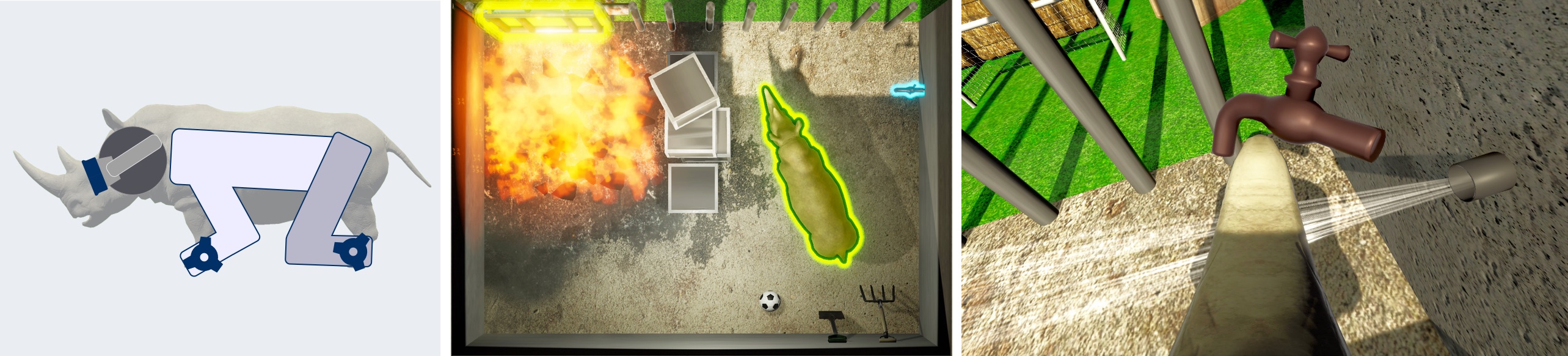}
\caption{
\textit{Rhino Room.} Players had to mimic the rhino posture (left) and escape from a burning zoo. The blue marked water tap (middle) had to be removed from the wall (right) to extinguish the fire. To open the yellow marked door, players had to use the horn and remove a lock bar (cf. \FG{fig:teaser2}).}
\label{fig:rhino}
\end{figure*}

\section{The Virtual Animal Experience}

Our main goal is to understand the benefits and limitations of animal avatars in VR games. Unfortunately, prior work indicates that we cannot overgeneralize such research, because animals vary greatly among themselves, be it regarding their skeletons, or postures, or motion. Hence, we focus on a sound methodology for a few sufficiently distinct representatives and provide in-depth insights how such avatars can be embedded in a gaming context. In particular, this section describes our reasoning regarding the choice of animals and their controls, as well as a quantitative evaluation of the outcomes.

\subsection{Choosing Virtual Animals}

One of the main questions to be asked when designing a game with nonhumanoid avatars is which creature to pick. Obviously, this choice is determined by various game design aspects that are not specific to animal avatars. However, the inclusion of such creatures adds degrees of freedom that need to be considered. We focus on two main aspects: the increased interaction design space and the induced challenges in controlling such avatars.

In the first place, playing an animal allows us to naturally inhibit the respective superhuman skills, such as flying as a bird or exploring underwater scenes as a dolphin. We postulate that such natural interactions could be intuitive and easy to learn when done right. Furthermore, the IVBO effect can intensify~\cite{banakou2013illusory,lugrin2016avatar} our perception of such actions due to the increased bond to our avatar compared to non-VR games.

These additional skills are often bound to additional body parts of nonhumanoid creatures. Fortunately, prior work~\shortcite{won2015homuncular} indicates that such additions do not necessarily destroy the IVBO effect and can still be intuitively controlled by players~\shortcite{Egeberg:2016:EHB:2927929.2927940}. In this respect, we recommend designing the avatar such that the altered morphology is perceived as an extension to our body, instead of being a restriction. For instance, Krekhov et al.~\shortcite{krekhov2018anim} reported that players liked the large wings of a bat because they felt like arm extensions and helpful tools, but disliked tiger paws that felt shorter than their actual limbs.

A second important aspect to be considered when designing such games is how the creature should be controlled by the player. To embody animal avatars, it is reasonable to synchronize the movements of the players as precisely as possible with their virtual representation~\cite{sanchez2010virtual}. However, typical room-scale VR equipment tracks only the players' heads and hands. We see three approaches to overcome that barrier: relying on only three tracked positions, including markerless tracking~\cite{8643070}, or requiring tracking extensions, such as the HTC Vive Trackers~\cite{vive}.

Even when full-body tracking is available, the question still remains how animals with significantly different postures should be controlled. A prominent example is creatures with non-upright postures, such as typical mammals. A straightforward way would be to crawl on all fours as a player to achieve the most realistic mapping. However, this might cause exhaustion over a longer period of time. As a remedy, \textit{half-body controls}~\shortcite{krekhov2018anim} can be applied to remain in an upright posture without noticeable sacrifice of IVBO. Half-body mapping approaches have either no direct mapping between players' legs and the limbs of an animal at all, or one leg is mapped to multiple limbs, which allows us to control creatures like spiders. Apart from fatigue, such controls are beneficial for cases where full-body tracking is not available.

To summarize, finding an optimal virtual creature is a multifaceted process,
and we suggest the impact of additional body parts and resulting superhuman skills, as well as possible control approaches be considered during the decision-making. To illustrate that process in more detail, we will showcase a possible selection approach of avatars and game mechanics in the next section.

\subsection{Example Realization}

To study animal avatars in a game context, we created a diverse testbed that supports multiple creatures with Unity 3D~\cite{unity}. The main idea is based on so-called \textit{escape rooms}~\cite{wiemker2015escape}: players are placed in a room filled with challenges that have to be solved in order to win/escape. We picked that setting for two particular reasons. First, if virtual and real rooms match in size and shape, locomotion can be achieved by natural walking, which has a positive impact on presence~\cite{ruddle2009benefits,Krekhov:2018:GVRA} and removes the need for additional, artificial navigation techniques, such as teleportation. Hence, players can focus more on the actual animal experience and are less distracted by accompanying functionalities. Second, escape room games are similar in their concept, which allows us to implement multiple, yet comparable scenarios, i.e., different rooms with different animal avatars. Each room contained two to three quests that involve navigation and object manipulation. In contrast to common escape games, we did not impose any time limitation to remove competition as a factor from our studies.

After picking the overarching game type, we focused on the design of the underlying game mechanics. We set ourselves the objective of building the individual room quests around distinct animal abilities. We selected three animals based their superhuman skills and/or additional body parts: a rhino, a scorpion, and a bird. In particular, our selection included morphologies with different degrees of similarity compared to our human body. A bird has a straightforward mapping, i.e., our arms become wings, and our legs become bird's feet. The horn of the rhino has no direct counterpart and requires a head-oriented interaction that is exotic for human beings. Finally, the scorpion comes with additional limbs, a tail with a sting, and two claws, which is the most differing morphology with at least two nonhumanoid interactions.

\begin{figure*}[t!]
\centering
\includegraphics[width=2.1\columnwidth]{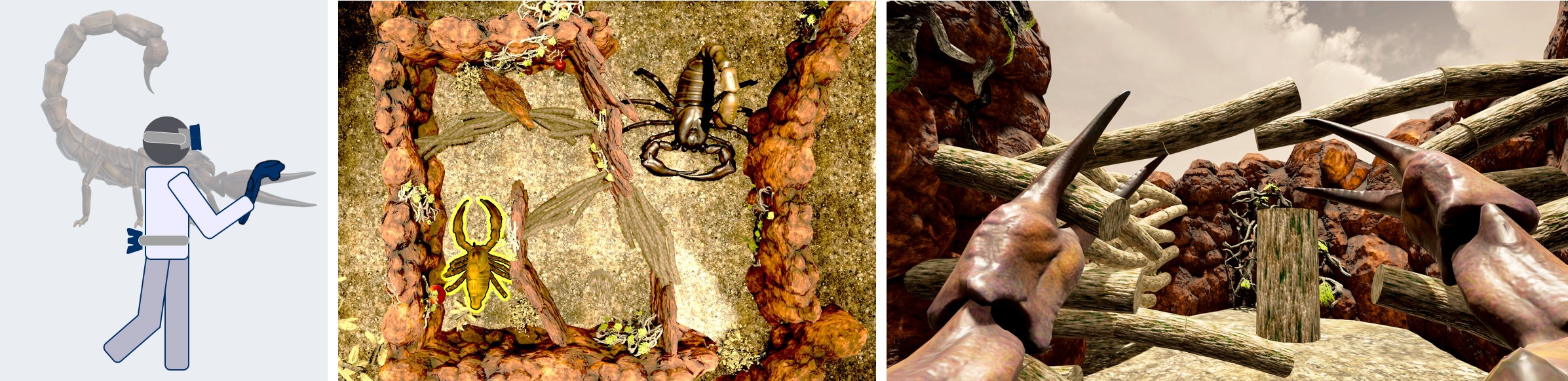}
\caption{
\textit{Scorpion Room.} Players remained in an upright posture (left) and used the controllers to open and close the claws and initiate a tail strike. To escape from the labyrinth, players had to cut away several branches (right). The exit-blocking emperor scorpion (middle) had to be pelted with poisoned fruits. The avatar tail was used to pick up these fruits. Aiming during the throwing process was done via a proper hip orientation.}
\label{fig:scorpion}
\end{figure*}

\subsubsection{Rhino Room}
We chose a rhino mainly because of its horn and the related capabilities. We suggest that such head-centered interactions occur seldom in VR games and could offer a unique player experience. In our case, players should use the horn (and paws) to escape from a burning cage, as shown in \FG{fig:rhino}. In particular, the horn was needed to move crates and clear the area in front of a water tap, to remove the tap from the wall to release a jet of water, and, finally, to lift and remove a lock bar that kept the door closed.

We utilized full-body controls with 1:1 motion mapping, i.e., players had to crawl on all fours during the game. Therefore, we positioned additional trackers at the hip and both ankles and wrists, i.e., no Vive controllers were used. We relied on inverse kinematics (IK)~\cite{buss2004introduction} to reconstruct the player posture. In particular, we applied a combination of closed-form and iterative solvers to provide the required degrees of freedom yet minimize jittering caused by unavoidable tracking errors. The horn was always visible to the players and placed slightly below the camera, as can be seen in \FG{fig:rhino}.

\subsubsection{Scorpion Room}
A scorpion offers even more unique interactions compared with a rhino if we allow players to control its tail and claws. In our scenario, depicted in in \FG{fig:scorpion}, players had to use these techniques to cut their way through a labyrinth and defeat a giant enemy by throwing a poisoned fruit at it (cf. \FG{fig:teaser2}). The fruit had to be picked up and thrown via the sting at the end of the scorpion tail.

To explore a variety of control approaches, we relied on half-body tracking, i.e., an upright posture, instead of 1:1 mapping, as done in the rhino game. We used the tracking data from the HMD, two Vive controllers, and an additional tracker at the hip position. Player arm movement was transferred to the virtual claws via IK. Trigger buttons could be used to open and close the claws to perform cutting. The circular track pad button initiated a tail strike, whereas aiming was performed by hip alignment. We did not track players' legs. Hence, the limbs of the scorpion were equipped with predefined ``walking'' animations matching the speed of player movement.

\subsubsection{Bird Cage}
To complete the diverse set of our virtual animals, we also included a flying creature, as can be seen in \FG{fig:bird}. Being a bird, players could use their virtual wings for two purposes: flying and creating gusts of wind to move objects. To escape from their virtual cage, players had to gain altitude, reach the highest point, and flutter with their wings in sync with the movement of a ceiling lamp, which then gained momentum and broke the cage door. Gaining altitude required significant effort, and players had to rest on rods between their flights.

We used the same tracking setup as in the rhino game, i.e., trackers at hip, wrists, and ankles. Players remained in an upright posture, and their arms were mapped to the wings, i.e., flapping was achieved via rapid up and down arm movements. To create gusts, players moved their arms horizontally instead. Flying around in the cage consisted of two components: flapping to gain altitude, and walking to perform a horizontal transition. We explicitly enforced that horizontal physical movement to minimize cybersickness~\cite{laviola2000discussion} by reducing the cognitive mismatch between physical and visual feedback. If players stopped waving their arms mid-air, a ``falling'' procedure was applied. That transition was performed rapidly to prevent cybersickness~\cite{Krekhov:2018:GVRA}.

\section{Evaluation}

\subsection{Research Questions and Hypotheses} 
The main purpose of our study was to investigate how players experience the animal avatars in our three game scenarios to draw conclusions about which aspects of representation, control, and interaction are perceived positively or negatively. Accordingly, our main research questions are: 
\begin{enumerate}[leftmargin=1cm]
\item[1:] Do animal avatars induce positive player experiences?
\item[2:] How do players evaluate the different design decisions regarding posture, visible body parts, and control mapping in our three games? 
\end{enumerate}

We assume that slipping into the role of an animal is a novel and interesting experience, and that the control of non-humanoid body parts and the use of related special abilities can raise players' enjoyment and engagement. Our three different realizations allow us to investigate whether a realistic posture and locomotion technique (e.g., crawling), the visibility of certain body parts, and the type of control mapping contribute to or interfere with a positive experience.  


Besides the general acceptance of animal avatars and the evaluation of the respective player experiences, we also consider the concept of IVBO. 
Based on prior findings indicating that IVBO is not limited to human-like bodies~\cite{won2015homuncular,krekhov2018anim},  
we hypothesize that our virtual animal bodies are capable of inducing IVBO as well, and that higher IVBO can be associated with higher perceived presence and game enjoyment. Hence, we want to test the following hypothesis: 

\begin{enumerate} [leftmargin=1cm]
\item[H1:] IVBO is positively correlated with game enjoyment and perceived presence. 
\end{enumerate}


\subsection{Study Procedure and Applied Measures}

We applied a within-subjects design with the game scenario as the independent variable with three levels (rhino, scorpion, bird). After being informed about the study procedure and signing an informed consent, participants filled in a first questionnaire about their demographic data, gaming habits, and prior experiences with VR headsets. We also administered the Immersive Tendencies Questionnaire (ITQ)~\cite{witmer1998measuring} to check participants' individual tendencies to get immersed in an activity or fiction. 

We then introduced the participants to the first game scenario. The three games were played in varying order to avoid biases due to sequence effects. In particular, we counterbalanced the sequence of the three game scenarios across subjects. All sessions followed the same procedure. First, the examiner explained the goal and controls of the game and applied the VR headset, an HTC Vive Pro~\cite{vive1} with a wireless adapter, and HTC Vive Trackers~\cite{vive}. Subsequently, a neutral mirror scene was started, in which participants saw their animal body avatar and were able to get used to the controls by observing their movements in a big mirror, as can be seen in \FG{fig:mirror}. This scene was displayed for 2 minutes to enable embodiment. The duration is a common choice for IVBO studies, and prior work indicates that even 15 seconds are enough to cause that effect~\cite{lloyd2007spatial}. After the mirror scene, we asked the participants to remove the HMD and administered the acceptance and control subscales of the alpha IVBO questionnaire~\cite{roth2017alpha}. We were mainly interested in the IVBO experience while playing and not in the subsequent effects on players' bodily perception, so the change dimension of the IVBO questionnaire was not applied. We decided against performing threat tests for capturing IVBO, because we expected significant sequence effects. Note there is no unified procedure for measuring IVBO, and a threat test is not the only possibility~\cite{kilteni2012sense,roth2017alpha}. We decided to use the alpha IVBO questionnaire and checked its reliability by calculating Cronbach's alpha for both subscales (all alphas > 0.82).

Upon completion, we re-equipped the participants with the HMD and launched the main game. Each gaming session lasted about 7 to 10 minutes, depending on how quickly players were able to solve all riddles. 
After each session, we asked the participants to fill in a questionnaire asking about their experiences during play. We administered the enjoyment subscale of the Intrinsic Motivation Inventory (IMI)~\cite{ryan2000self} to assess general game enjoyment, as well as the Player Experience of Need Satisfaction (PENS) questionnaire~\cite{Ryan.2006, Rigby.2007, Johnson.2010} to test experienced autonomy, competence, and intuitiveness of controls. We measured the feelings of presence by the Presence Questionnaire~\cite{Witmer.1998,Witmer.2005} and the Igroup Presence Questionnaire (IPQ)~\cite{Schubert.2003, Schubert.2018}. 
To test for negative physiological effects of using the immersive HMD, we also administered the Simulator Sickness Questionnaire (SSQ)~\cite{kennedy1993simulator}. Finally, we posed some custom, game-specific questions to assess how players evaluated the controls, the required posture during play, as well as the visibility of certain body parts. We also asked whether participants could imagine using this kind of avatar control in other VR games. All administered questionnaire items had to be rated on a unipolar scale ranging from 0 to 6 (``completely disagree'' to ``completely agree''), except from the SSQ, which had to be rated on a unipolar 4-point scale.

\begin{figure}[t!]
\centering
\includegraphics[width=1.0\columnwidth]{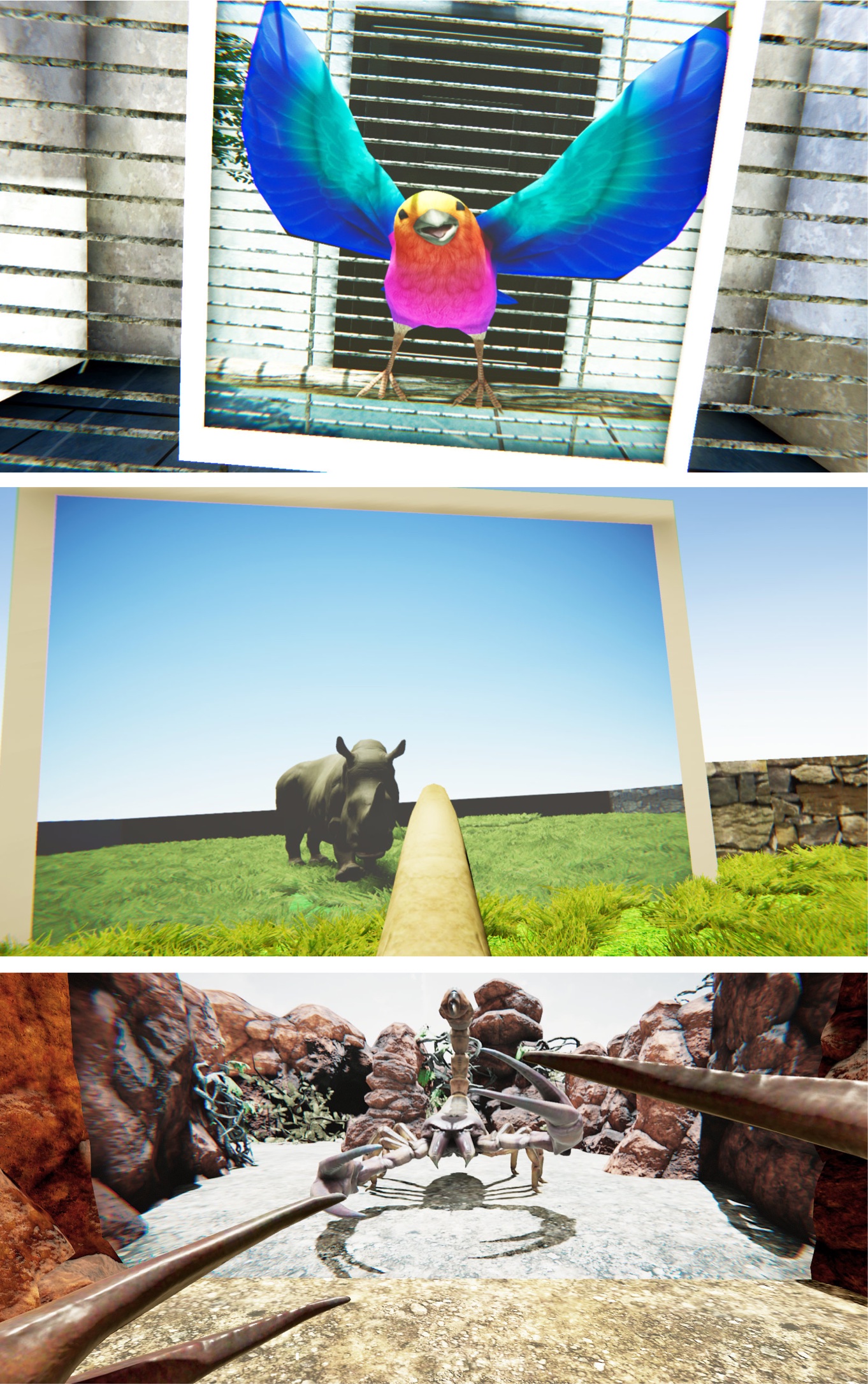}
\caption{
Before each game, players were asked to act in front of a wall-sized mirror for about two minutes to get familiar with their virtual representation and to answer the alpha IVBO questionnaire~\protect\cite{roth2017alpha}.}
\label{fig:mirror}
\end{figure}

\begin{table*}[h]

  \begin{tabularx}{\textwidth} {>{\raggedright\arraybackslash}p{0.7cm} >{\raggedright\arraybackslash}p{4.5cm}>{\centering\arraybackslash}X>{\centering\arraybackslash}X >{\centering\arraybackslash}X>{\centering\arraybackslash}X>{\centering\arraybackslash}X>{\centering\arraybackslash}X}
    \toprule
    \addlinespace
   &  & \textbf{Rhino} & \textbf{Scorpion}  & \textbf{Bird}  \\
\addlinespace     
   \multicolumn{2}{l}{} & \textit{M} (\textit{SD}) & \textit{M} (\textit{SD}) & \textit{M} (\textit{SD})  & \textit{F} & {$\chi^2$} & \textit{ p} \\
    \midrule
    \addlinespace\addlinespace
    \textbf{IMI}
    & Enjoyment/Interest        &  4.46 (0.96)  & 4.17 (1.37) & 4.08 (1.36) & - & 0.065 & .968 \phantom{*}\\
\addlinespace\addlinespace    
    \textbf{PENS} 
    & Competence            &  3.66 (1.85)  & 3.03 (1.85) & 3.33 (1.66) & - & 3.000 &  .223 \phantom{*}\\
    & Autonomy              &    3.82 (1.51)  & 3.27 (1.53) & 3.14 (1.62) & 3.562 & - & .034 * \\ 
    & Intuitive controls        &  5.14 (1.15)  & 4.50 (1.74) & 4.53 (1.35) & - &   11.608 & .003 *\\
    \addlinespace\addlinespace
    \textbf{PQ} 
    & Realism               &  4.09 (1.15)  & 3.93 (1.19) & 3.92 (1.09) & 0.493 & - & .613 \phantom{*}\\
    & Possibility to act        &  4.55 (0.83)  & 3.98 (0.99) & 3.85 (1.06) & 7.607 & - & .001 *  \\
    & Quality of interface        &  4.91 (1.02)  & 4.54 (1.21) & 4.97 (1.02) & 4.142 & - & .020 *\\
    & Possibility to examine      &  4.44 (1.05)  & 3.89 (1.29) & 3.72 (1.24) &-  & 22.709 & < .001 *\\
    & Self-evaluation of performance  &  4.53 (1.51)  & 4.19 (1.44) & 4.34 (1.24) &-  & 4.019 & .134 \phantom{*}\\
    & Total               &  4.42 (0.92)  & 4.06 (1.02) & 4.09 (0.91) & 4.162 & - & .020 *\\
    \addlinespace\addlinespace
    \textbf{IPQ} 
    & General               &  3.88 (1.52)  & 3.56 (1.68) & 3.63 (1.70) &-  & 2.092 & .351 \phantom{*}\\
    & Spatial presence          &  4.22 (1.19)  & 4.07 (1.24) & 3.94 (0.99) & 1.339 &- & .270 \phantom{*}\\
    & Involvement             &  3.34 (1.28)  & 3.02 (1.21) & 2.96 (1.26) & 2.373 &- & .102 \phantom{*}\\
    \addlinespace\addlinespace
    \textbf{IVBO} 
    & Acceptance            &  3.34 (1.15)  & 3.26 (1.29) & 3.46 (1.31) & 0.502 &- & .608 \phantom{*} \\
    & Control               &  4.64 (1.42)  & 4.40 (1.53) & 4.88 (1.26) &-  & 3.576 & .167 \phantom{*}\\
  \addlinespace\addlinespace
  \textbf{SSQ} 
  & Nausea              &  0.21 (0.32)  & 0.27 (0.38) & 0.26 (0.40) &-  & 2.590 & .274 \phantom{*}\\
  & Oculomotor            &  0.24 (0.27)  & 0.35 (0.45) & 0.26 (0.34) &-  & 9.968 & .007 *\\
  & Disorientation          &  0.11 (0.18)  & 0.19 (0.41) & 0.17 (0.42) &-  & 3.405 & .182 \phantom{*}\\
  \bottomrule
\end{tabularx}
\begin{tablenotes}\footnotesize \raggedleft\arraybackslash
        \item * significant main effect at a significance level of $\alpha$ = .05
      \end{tablenotes}
\caption{Mean scores and standard deviations of the IMI, PENS, PQ, IPQ, IVBO, and SSQ subscales for the three game scenarios (all scales range from 0 to 6, except from SSQ, which ranges from 0 to 3). Significant differences of mean values between conditions were tested by calculating repeated measures ANOVA (\textit{F}) or Friedman tests ($\chi^2$), if data was not normally distributed.} 
\label{tab:Animals_Means}
\end{table*}

\subsection{Results}

In total, 32 persons (19 female, 13 male) with a mean age of 23.7 years (\textit{SD}~=~5.18) participated in our study. Due to recruiting at a university, most of them were students (\textit{N}~=~25), whereas the others were employees. Many participants reported prior experiences with VR headsets (\textit{N}~=~22), but only two of them used VR systems on a regular basis. All participants were familiar with digital games and the majority (\textit{N}~=~24) reported playing digital games regularly.


\subsubsection{Players' Experiences with the Three Animal Avatars} 
Following our research questions, we analyzed participants' ratings of the different animal avatars and their experiences in the three game scenarios.
Mean values of all applied questionnaires can be found in \TA{tab:Animals_Means}. 
Considering the scales' range from 0 to 6, almost all aspects were rated above average, indicating a positive experience in all three game scenarios. In particular, IMI scores and perceived presence as measured by the PQ show that players enjoyed the games and felt as if they were actually being and acting in the virtual world.
Scores of all three subscales of the SSQ---nausea, oculomotor, and disorientation---were very low in all conditions (all \textit{M} < 0.36), and thus cybersickness was not an issue and can be excluded as a potential confounding variable.

We compared players' experiences in the three game scenarios in terms of the subscales of IMI, PENS, PQ, IPQ, and IVBO to investigate whether the different avatars and interactions were perceived differently. Our analysis of covariance indicated no significant influence of immersive tendencies (ITQ) on our dependent variables, hence we did not further elaborate on that. In advance, we performed Kolmogorov-Smirnov tests to assess all scales for normal distribution as a requirement for parametric calculations. If violated, results of Friedman tests are reported instead of repeated measures ANOVA for comparing the three game scenarios. Bonferroni correction was applied for all post-hoc tests. The main test statistics can be found in \TA{tab:Animals_Means}. 

Regarding players' need satisfaction (PENS), we found significant differences between the three game scenarios in terms of autonomy and intuitive controls. Post-hoc tests indicate that perceived autonomy was significantly higher when playing with the rhino compared to the scorpion scenario (\textit{p}~=~.049). 
The intuitiveness of controls was rated significantly higher in the rhino scenario than in both the scorpion (\textit{p}~=~.012) and the bird condition (\textit{p}~=~.004). 

Comparisons of the PQ subscales show further significant differences. According to post-hoc tests, participants reported significantly better experiences regarding the possibility to act and the possibility to examine in the rhino scenario than in the other two games (all \textit{p}~<~.004). 
Moreover, the total score for presence was significantly higher for the rhino than for the scorpion (\textit{p}~=~.017). 
The interface quality, in contrast, was rated significantly lower in the scorpion game compared to the bird scenario (\textit{p}~=~.049). 
All other measures did not differ significantly, i.e., we did not find significant differences regarding general game enjoyment or IVBO.

\def\rot{\rotatebox}

\begin{table*}[h]

  \begin{tabularx}{\textwidth} {>{\raggedright\arraybackslash}p{0.1cm} >{\raggedright\arraybackslash}p{2cm}>{\centering\arraybackslash}X>{\centering\arraybackslash}X >{\centering\arraybackslash}X>{\centering\arraybackslash}X>{\centering\arraybackslash}X >{\centering\arraybackslash}X >{\centering\arraybackslash}X >{\centering\arraybackslash}X >{\centering\arraybackslash}X>{\centering\arraybackslash}X>{\centering\arraybackslash}X>{\centering\arraybackslash}X>{\centering\arraybackslash}X}
    \toprule
    \addlinespace   \addlinespace 
  &  & \textbf{IMI}  & \multicolumn{6}{c}{\textbf{PQ}} & \multicolumn{3}{c} {\textbf{IPQ}} \\
  \addlinespace   
  \cmidrule(lr){3-3}   \cmidrule(lr){4-9} \cmidrule(lr){10-12} 
    \addlinespace 
    &  & \rot{45}{enjoyment} & \rot{45}{realism} & \rot{45}{\parbox{2cm}{possibility \\ \phantom{lol}to act}} & \rot{45}{interface quality}  & \rot{45}{\parbox{2cm}{possibility \\ \phantom{lol}to examine}} &  \rot{45}{performance} & \rot{45}{total} &  \rot{45}{general} & \rot{45}{spatial presence} & \rot{45}{involvement} \\
    
\addlinespace     
   \multicolumn{2}{l}{} & \textit{r\textsubscript{s}} (\textit{p})  & \textit{r\textsubscript{s}} (\textit{p}) & \textit{r\textsubscript{s}} (\textit{p}) & \textit{r\textsubscript{s}} (\textit{p}) & \textit{r\textsubscript{s}} (\textit{p}) & \textit{r\textsubscript{s}} (\textit{p}) & \textit{r\textsubscript{s}} (\textit{p})& \textit{r\textsubscript{s}} (\textit{p})& \textit{r\textsubscript{s}} (\textit{p})& \textit{r\textsubscript{s}} (\textit{p})
   \\
    \midrule
    \addlinespace \addlinespace 
    \multicolumn{2}{l}{\textbf{IVBO - Rhino}} \\\addlinespace
    & Acceptance            &  0.068 (.710) &  0.511* (.003)  & 0.312* (.049) & 0.480* (.005) & 0.404* (.022) & 0.329 (.066) & 0.508* (.003)  & 0.192 (0.291) & 0.354* (.047) & 0.046 (.805)\\
    \addlinespace & Control               & 0.442* (.011) & 0.734* (<.001)  & 0.557* (.001) & 0.689* (<.001)  & 0.612* (<.001) & 0.467* (.007) & 0.760* (<.001) & 0.559* (.001) & 0.590* (<.001) & 0.022 (.903) \\
\addlinespace \addlinespace 
    \multicolumn{2}{l}{\textbf{IVBO - Scorpion}} \\\addlinespace
    & Acceptance            &  0.357* (.045)  & 0.428* (.015) & 0.438* (.012) & 0.272 (.132)  & 0.461* (.008) & 0.399* (.024) & 0.460* (.008) & 0.560* (.001) & 0.496* (.004) & 0.349 (.051)  \\
    \addlinespace & Control               &  0.572* (.001)  & 0.742* (<.001)  & 0.563* (.001)   & 0.562* (.001) & 0.642* (<.001)  & 0.586* (<.001) & 0.769* (<.001) & 0.460* (.008) & 0.504* (.003) & 0.190 (0.297)\\
\addlinespace \addlinespace 
    \multicolumn{2}{l}{\textbf{IVBO - Bird}} \\\addlinespace
    & Acceptance            & 0.397* (.024) & 0.481* (.005) & 0.585* (<.001)  & 0.451* (.010)&  0.557* (.001) & 0.391* (.027) & 0.618* (<.001) & 0.436* (.013) & 0.480* (.005) & 0.237 (.192)\\
    \addlinespace & Control               & 0.387* (.029) & 0.358* (.044) & 0.412* (.019) & 0.591* (<.001) &  0.478* (.006) & 0.324 (.071) & 0.501* (.004) & 0.474* (.006) & 0.520* (.002) & 0.163 (.374)\\
  \bottomrule
\end{tabularx}
\begin{tablenotes}\footnotesize \raggedleft\arraybackslash
        \item * significant correlation at a significance level of $\alpha$ = .05
      \end{tablenotes}
 \caption{Spearman's rank-order correlation coefficients \textit{r\textsubscript{s}} and \textit{p}-values that indicate correlations among the IVBO subscales and IMI, PQ, and IPQ.} 
 \label{tab:correlations}
\end{table*}

\subsubsection{Insights About the Different Postures}
As our different scenarios required different postures, we asked how the actual gaming posture was perceived and if participants would have preferred another posture. For the bird and scorpion avatars, participants agreed that the upright posture was comfortable (bird: \textit{M}~=~4.50, \textit{SD}~=~1.48; scorpion: \textit{M}~=~5.03, \textit{SD}~=~1.26), whereas the kneeling posture in the rhino condition was rated ambiguously and perceived as being physically demanding by several participants (\textit{M}~=~3.44, \textit{SD}~=~1.98). However, when asked whether they would prefer an upright playing posture to control the rhino, the majority of players tended to disagree (\textit{M}~=~2.56, \textit{SD}~=~2.41). 
In contrast, they agreed that the kneeling posture contributed to the realism of the game (\textit{M}~=~4.16, \textit{SD}~=~1.99). 

The bird posture and the mechanics of locomotion (flapping with the arms to move up combined with walking to move horizontally) was also perceived as being realistic (\textit{M}~=~4.84, \textit{SD}~=~1.08). 
Accordingly, participants did not wish for another posture (\textit{M}~=~1.09, \textit{SD}~=~1.65). 

Similar ratings were given for the scorpion: participants rated the posture as being realistic (\textit{M}~=~4.13, \textit{SD}~=~1.95) and did not wish for another posture such as kneeling (\textit{M}~=~1.78, \textit{SD}~=~2.01), 
although a kneeling posture would be objectively more realistic. When asked whether they had the feeling of being stuck in the ground (due to the low head position), participants were rather inconclusive (\textit{M}~=~3.25, \textit{SD}~=~2.17). 
During the experiment, we observed that some participants were indeed a bit irritated at the beginning, but got used to the mismatch between their own and the avatar's body size quite quickly. 

\subsubsection{Controls}
Overall, the high ratings for PENS' intuitive controls confirm that participants had no problems moving and interacting in the game world and using the animals' abilities in all three scenarios.
Although our three animal avatars are rather different in terms of posture and control mapping, participants stated in all three cases that they could very well imagine using this kind of avatar control in other VR games (rhino: \textit{M}~=~4.56, \textit{SD}~=~1.78; scorpion: \textit{M}~=~4.94, \textit{SD}~=~1.44; bird: \textit{M}~=~4.75, \textit{SD}~=~1.55).


\subsubsection{Visibility of Body Parts}
Regarding the visibility of certain body parts, we were interested in players' opinions about the usefulness of such visualizations and the possible interferences. In the rhino game, the horn was displayed in the players' sight throughout the game. However, the horn was neither perceived as being disruptive (\textit{M}~=~0.69, \textit{SD}~=~1.18) 
nor resulted in the perception of a constrained field of view (\textit{M}~=~0.97, \textit{SD}~=~1.26). 
In contrast, participants enjoyed using the horn as a tool (\textit{M}~=~4.78, \textit{SD}~=~1.60). 

In the bird cage scenario, apart from the wings, the bird's feet were also displayed. Participants appreciated this display, because they rated this feature as being helpful for landing on the rods (\textit{M}~=~5.22, \textit{SD}~=~1.49). 

 \subsubsection{Special Abilities}

We also asked players about their opinions regarding the special abilities they could use as animals. Participants agreed that the use of the horn of the rhino enriched the whole experience (\textit{M}~=~4.91, \textit{SD}~=~1.33). 
The scorpion's sting was rated as very interesting (\textit{M}~=~4.44, \textit{SD}~=~1.59), 
and players also liked to use the claws (\textit{M}~=~3.81, \textit{SD}~=~1.86). 
Moreover, players rated the experience of flying as a bird as very interesting (\textit{M}~=~5.12, \textit{SD}~=~1.36) 
as they did the bird's ability to create gusts of wind (\textit{M}~=~4.19, \textit{SD}~=~1.93). 

\subsubsection{Correlations between IVBO, Enjoyment, and Presence}
Mean values of IVBO control are rather high, and mean values of IVBO acceptance are above average, as well, which indicates that players have experienced IVBO while controlling our animal avatars. 
To test our hypothesis regarding the relation between IVBO and the player experience (H1), we analyzed the correlations between the two subscales of the IVBO questionnaire and the subscales of IMI, PQ, and IPQ.
We calculated Spearman's rank correlation coefficients (Spearman’s rho) due to a lack of normal distribution of some scales. Table~\ref{tab:correlations} summarizes the results for each game scenario. 

Overall, we found significant positive correlations between IVBO and nearly all PQ and IPQ subscales: ratings of experienced realism, the possibility to act, the possibility to examine, PQ total, and spatial presence are consistently significantly correlated with both IVBO dimensions in all three games. 
Furthermore, IVBO control also significantly correlates with the perceived interface quality and the general feeling of presence as measured by the IPQ. IVBO acceptance correlates with the interface quality except from the scorpion scenario, and with general presence except from the rhino scenario. 
The only scale not significantly correlated with IVBO in any scenario is IPQ involvement. 

Regarding game enjoyment, our analysis shows significant positive correlations between IVBO control and IMI enjoyment scores in all three scenarios. The correlation between IVBO acceptance and enjoyment is significant in the bird cage and the scorpion room, whereas there is no correlation in the rhino condition.
In sum, our results mainly support our hypothesis H1.




\subsection{Discussion and Design Implications}

Our results indicate that animal avatars in VR games can induce positive player experiences. We implemented three games with animal avatars that are very different regarding body features and abilities, and in all cases players reported high enjoyment and high presence, i.e., the feeling of actually being in the virtual world and being the rhino, scorpion, or bird. Participants particularly appreciated the novel body experiences and nonhumanoid perspectives, as well as the use of the special animal abilities.

\subsubsection{Special abilities}
The feedback of participants on our three games shows that players are very interested in performing actions that they are not able to perform in real life. For instance, they were fascinated by the ability to fly upwards using their wings as a bird, and they enjoyed testing how they could manipulate objects with their rhino horn. 
We reason that such superhuman abilities significantly contribute to players' enjoyment and their motivation to play. Hence, the main game mechanics of games featuring animal avatars should \textbf{foster the animal's specific characteristics and abilities} to create novel, fanciful experiences. Designers should take advantage of players' curiosity and expose unique animal features.

\subsubsection{Player Posture and Controls}

In all three games, the adopted postures were perceived positively and without an explicit desire for alternatives. In other words, there is no indication that a realistic yet uncomfortable posture (rhino) is better or worse than a convenient and upright but unrealistic posture (scorpion). However, the statements regarding crawling on the floor as a rhino were quite ambivalent, i.e., some of the participants enjoyed such an experience, whereas others became rapidly exhausted by that activity. Note that the rhino, however, outperformed other animals in certain subscales, such as autonomy (PENS), intuitive controls (PENS), and the possibility to act (PQ). Hence, a 1:1 mapping where players have to behave exactly like they would expect from their animal avatar is easier to grasp and is perceived as very realistic.

Similarly, our results did not disqualify or favor any particular control approach -- all three controls were rated as very intuitive and participants could imagine using such approaches in other VR games. Hence, we suggest \textbf{controls be designed based on the game-related animal abilities and the target audience}. For instance, we assume that children are more willing to spend their time crawling on the floor compared to elderly adults. In general, transferring as many player movements as possible onto the avatar is a reasonable approach, especially considering the positive influence on IVBO~\cite{sanchez2010virtual}. However, as we have seen in the scorpion case, less straightforward mappings can be equally engaging and fun without enforcing an uncomfortable posture. Furthermore, such implementations can be achieved with less tracking equipment.

\subsubsection{Visible body parts}
Game designers have different approaches regarding the visibility of the avatar's body in first-person mode. From our experience, we would not recommend visualizing the whole body, as the avatar head position often leads to confusing viewports when players look down on them. Instead, we suggest \textbf{the visualization be limited to body parts that can be directly controlled by the player}, e.g., claws, tails, wings, and horns. In particular, the additional body parts, although reducing the visible area, are not perceived as disturbing. For instance, participants rated the horn as a helpful tool and reported that the bird's feet facilitate the landing on thin rods. Furthermore, seeing animal body parts like claws moving in sync with our own body increases our awareness of embodiment. 

\subsubsection{Morphology}
Considering the morphology of our three animal avatars, our results indicate that players had no problems with controlling bodies that are not similar to the human shape. Even the control of the scorpion, which has several additional limbs, claws, and a tail, was perceived as intuitive and did not cause any confusion.
In contrast, we observed that players particularly liked additional body parts such as the scorpion's sting or the horn of the rhino. 
Hence, we challenge game designers to \textbf{consider extraordinary animal shapes} and derive innovative game mechanics. We should not back off from adopting complex body compositions as long as they are associated with interesting possibilities for interaction design.
In our three games, we always focused on the outstanding bodily features of the animals and linked them to certain player abilities (e.g., create gusts of wind) to \textbf{give significance to them}. 
We suggest that additional or missing body parts compared to the human body should enrich players' opportunities to examine and interact with the virtual world and not appear as an impairment. This way, we can foster players' experience of having superhuman capabilities.

\subsubsection{IVBO}
Finally, we conclude that additional body parts or a nonhuman body shape do not inhibit an avatar's potential to induce IVBO. Our three exemplary animal avatars illustrate that IVBO is not limited to body models that are similar to the human body. With regard to our hypothesis H1, our results reveal that IVBO---which was measured prior to the gaming sessions and is, thus, not biased by the subsequent game experience---is positively correlated with game enjoyment and perceived presence. This finding indicates that IVBO may contribute to a positive player experience.
Hence, we conclude that \textbf{IVBO is a considerable factor when designing nonhumanoid VR avatars}.
To foster IVBO in a game, we suggest that game designers provide players with possibilities to see their virtual body (e.g., in mirrors or water reflections) to increase awareness of their virtual representation.

\section{Limitations of the Study}
Our derived design implications are based especially on the three evaluated scenarios. Hence, we need to consider a set of associated limitations to prevent possible misinterpretation of the findings. In the first place, our main goal was to expose a complete pipeline of embedding animal avatars into VR games. We aimed to raise the awareness regarding the wide variety of decisions (e.g., posture, animal type, mapping/controls, special abilities, morphologies, locomotion) that have to be considered during such a game development process. As a result, our evaluated scenarios are rather complex games with a number of possibly influential variables that might limit the generalizability. For instance, the general appeal of an animal, e.g., a dangerous scorpion vs. a domestic bird, might impact our game enjoyment. Well-known species, such as a rhino, might be more intuitive to control than exotic creatures with abilities unknown to us. And although we removed artificial VR navigation techniques (e.g., teleportation) by matching the size of the virtual environment to the physical room, the different locomotion (flying vs. crouching vs. walking) could still have a considerable impact on the player experience. Finally, although all three scenarios were escape games, the particular quests could have influenced the outcome. In other words, we emphasize that the direct comparison of the three study conditions should be interpreted with these limitations in mind and that the reason behind our variety of scenarios was not the comparison per se, but our strive to cover as much of the animal avatar design space as possible to create a comprehensive starting point for future explorations.  Also, comparative studies in the future would benefit from an additional control group with a human avatar for a better assessment of the animals’ influence on IVBO and game enjoyment compared to a rather traditional virtual representation.

Another important limitation to be mentioned is that we did not involve animal/domain experts during the design phase and pretests of our study. Our decision making regaring the choice of animal avatars and, even more important, their abilities and interactions, was made without the input of an expert. The latter could have provided additional input regarding the realistic behavior of animals and our perception of such species.

As a next step, we suggest to focus on particular avatar components in a more targeted study to build a theoretical framework that provides an isolated in-depth exploration of major factors, such as locomotion, altered or additional body parts, and appeal. We suggest that such isolated insights should be gathered as a second step after seeing the “whole picture”, i.e., how such animals work or do not work in games. For instance, prior work~\shortcite{krekhov2018anim} reported that embodying a tiger while crawling on all fours was disliked by the participants, whereas a rhino, being a very similar mammal, provided the highest enjoyment in our scenario. Hence, we suppose that it is not just the familiarity with an animal or the intuitive locomotion, but rather subtle details, e.g., the additional horn, that can significantly alter our experience of such avatars and, thus, need further research.

For further studies, we also recommend expanding the age range of the participants. In our case, most participants were students due to the acquisition at the university, which limits the applicability of our findings. Instead, it is likely that aspects such as the necessary physical effort or the perceived avatar appeal are experienced differently by other age groups. Consequently, the age of the target audience might be an important design consideration and should be explored in future work.

\section{Conclusion and Future Work}

Our work investigated the hidden potential of animal avatars. We focused on virtual reality games because of the related IVBO effect that allows us to embody our avatar and perceive certain player interactions in a more intensive way. Accordingly, our studies supported our general assumption that games created around animal avatars could lead to great enjoyment. In particular, players liked the interactions resulting from additional body parts, such as wings and horns. In this regard, we proposed different ways to control animals with such differing morphologies and discussed related design implications for animal-centered VR games.

As a particular finding, we reported a correlation among IVBO, presence, and game enjoyment. Since our studies had a different emphasis, i.e., the general usefulness of animal avatars, we cannot disentangle these relations in detail. However, we see our results as evidence for the importance of IVBO for VR games in general, be it human or animal avatars. Hence, we propose an in-depth investigation of that overarching topic as possible follow-up research. Ultimately, we assume that a further exploration will encourage researchers and practitioners to consider IVBO as a helpful tool that allows the creation of novel, engaging player experiences that cannot be realized in non-VR games.

\balance{}

\bibliographystyle{SIGCHI-Reference-Format}
\bibliography{beastly}

\end{document}


%% file: ivda-macros.tex
\usepackage{microtype}

\usepackage{ulem}
\makeatletter
\def\ifEmpty#1{\def\@temp{#1}\ifx\@temp\@empty}
\makeatother

\usepackage{xspace}

\usepackage{amsmath,amsfonts,amssymb}
\usepackage{url} 

\newcommand{\FG}[1]{Figure~\ref{#1}}

\newcommand{\TA}[1]{Table~\ref{#1}}

\newcommand{\shah}{{\textstyle \amalg{\kern-4.pt\amalg}}}

%% file: beastly.bbl

\begin{thebibliography}{00}


\ifx \showCODEN    \undefined \def \showCODEN     #1{\unskip}     \fi
\ifx \showDOI      \undefined \def \showDOI       #1{{\tt DOI:}\penalty0{#1}\ }
  \fi
\ifx \showISBNx    \undefined \def \showISBNx     #1{\unskip}     \fi
\ifx \showISBNxiii \undefined \def \showISBNxiii  #1{\unskip}     \fi
\ifx \showISSN     \undefined \def \showISSN      #1{\unskip}     \fi
\ifx \showLCCN     \undefined \def \showLCCN      #1{\unskip}     \fi
\ifx \shownote     \undefined \def \shownote      #1{#1}          \fi
\ifx \showarticletitle \undefined \def \showarticletitle #1{#1}   \fi
\ifx \showURL      \undefined \def \showURL       #1{#1}          \fi

\bibitem{ahn2016experiencing}
{Sun~Joo Ahn}, {Joshua Bostick}, {Elise Ogle}, {Kristine~L Nowak}, {Kara~T
  McGillicuddy}, {and} {Jeremy~N Bailenson}. 2016.
\newblock \showarticletitle{Experiencing nature: Embodying animals in immersive
  virtual environments increases inclusion of nature in self and involvement
  with nature}.
\newblock {\em Journal of Computer-Mediated Communication\/} {21}, 6 (2016),
  399--419.
\newblock


\bibitem{bach2003sensory}
{Paul Bach-y Rita} {and} {Stephen~W Kercel}. 2003.
\newblock \showarticletitle{Sensory substitution and the human--machine
  interface}.
\newblock {\em Trends in cognitive sciences\/} {7}, 12 (2003), 541--546.
\newblock


\bibitem{banakou2013illusory}
{Domna Banakou}, {Raphaela Groten}, {and} {Mel Slater}. 2013.
\newblock \showarticletitle{Illusory ownership of a virtual child body causes
  overestimation of object sizes and implicit attitude changes}.
\newblock {\em Proceedings of the National Academy of Sciences\/} {110}, 31
  (2013), 12846--12851.
\newblock


\bibitem{berenguer2007effect}
{Jaime Berenguer}. 2007.
\newblock \showarticletitle{The effect of empathy in proenvironmental attitudes
  and behaviors}.
\newblock {\em Environment and Behavior\/} {39}, 2 (2007), 269--283.
\newblock


\bibitem{Biocca:1995:IVR:207922.207926}
{Frank Biocca} {and} {Ben Delaney}. 1995.
\newblock \showarticletitle{Communication in the Age of Virtual Reality}.
\newblock L. Erlbaum Associates Inc., Hillsdale, NJ, USA, Chapter Immersive
  Virtual Reality Technology, 57--124.
\newblock
\showISBNx{0-8058-1550-3}
\showURL{%
\url{http://dl.acm.org/citation.cfm?id=207922.207926}}


\bibitem{blom2014effects}
{Kristopher~J Blom}, {Jorge Arroyo-Palacios}, {and} {Mel Slater}. 2014.
\newblock \showarticletitle{The effects of rotating the self out of the body in
  the full virtual body ownership illusion}.
\newblock {\em Perception\/} {43}, 4 (2014), 275--294.
\newblock


\bibitem{botvinick1998rubber}
{Matthew Botvinick} {and} {Jonathan Cohen}. 1998.
\newblock \showarticletitle{Rubber hands ‘feel’touch that eyes see}.
\newblock {\em Nature\/} {391}, 6669 (1998), 756.
\newblock


\bibitem{buss2004introduction}
{Samuel~R Buss}. 2004.
\newblock \showarticletitle{Introduction to inverse kinematics with jacobian
  transpose, pseudoinverse and damped least squares methods}.
\newblock {\em IEEE Journal of Robotics and Automation\/} {17}, 1-19 (2004),
  16.
\newblock


\bibitem{cairns2014immersion}
{Paul Cairns}, {Anna Cox}, {and} {A~Imran Nordin}. 2014.
\newblock \showarticletitle{Immersion in digital games: review of gaming
  experience research}.
\newblock {\em Handbook of digital games\/} (2014), 337--361.
\newblock


\bibitem{vive1}
{HTC Corporation}. 2018.
\newblock {HTC Vive}.
\newblock Website.   (2018).
\newblock
\newblock
\shownote{Retrieved October 12, 2018 from \url{https://www.vive.com/}.}


\bibitem{vive}
{HTC Corporation}. 2019.
\newblock {HTC Vive Tracker}.
\newblock Website.   (2019).
\newblock
\newblock
\shownote{Retrieved March 10, 2019 from
  \url{https://www.vive.com/eu/vive-tracker/}.}


\bibitem{Egeberg:2016:EHB:2927929.2927940}
{Mie C.~S. Egeberg}, {Stine L.~R. Lind}, {Sule Serubugo}, {Denisa Skantarova},
  {and} {Martin Kraus}. 2016.
\newblock \showarticletitle{Extending the Human Body in Virtual Reality: Effect
  of Sensory Feedback on Agency and Ownership of Virtual Wings}. In {\em
  Proceedings of the 2016 Virtual Reality International Conference} {\em (VRIC
  '16)}. ACM, New York, NY, USA, Article 30, 4 pages.
\newblock
\showISBNx{978-1-4503-4180-6}
\showDOI{%
\url{http://dx.doi.org/10.1145/2927929.2927940}}


\bibitem{ehrsson2007experimental}
{H~Henrik Ehrsson}. 2007.
\newblock \showarticletitle{The experimental induction of out-of-body
  experiences}.
\newblock {\em Science\/} {317}, 5841 (2007), 1048--1048.
\newblock


\bibitem{ehrsson2009many}
{H~Henrik Ehrsson}. 2009.
\newblock \showarticletitle{How many arms make a pair? Perceptual illusion of
  having an additional limb}.
\newblock {\em Perception\/} {38}, 2 (2009), 310--312.
\newblock


\bibitem{guterstam2011illusion}
{Arvid Guterstam}, {Valeria~I Petkova}, {and} {H~Henrik Ehrsson}. 2011.
\newblock \showarticletitle{The illusion of owning a third arm}.
\newblock {\em PloS one\/} {6}, 2 (2011), e17208.
\newblock


\bibitem{heeter1992being}
{Carrie Heeter}. 1992.
\newblock \showarticletitle{Being there: The subjective experience of
  presence}.
\newblock {\em Presence: Teleoperators \& Virtual Environments\/} {1}, 2
  (1992), 262--271.
\newblock


\bibitem{IJsselsteijn}
{Wijnand~A. IJsselsteijn}, {Huib de Ridder}, {Jonathan Freeman}, {and}
  {Steve~E. Avons}. 2000.
\newblock Presence: concept, determinants, and measurement.
\newblock   (2000).
\newblock
\showDOI{%
\url{http://dx.doi.org/10.1117/12.387188}}


\bibitem{jo2017impact}
{Dongsik Jo}, {Kangsoo Kim}, {Gregory~F Welch}, {Woojin Jeon}, {Yongwan Kim},
  {Ki-Hong Kim}, {and} {Gerard~Jounghyun Kim}. 2017.
\newblock \showarticletitle{The impact of avatar-owner visual similarity on
  body ownership in immersive virtual reality}. In {\em Proceedings of the 23rd
  ACM Symposium on Virtual Reality Software and Technology}. ACM, 77.
\newblock


\bibitem{Johnson.2010}
{Daniel Johnson} {and} {John Gardner}. 2010.
\newblock \showarticletitle{Personality, motivation and video games}. In {\em
  Proceedings of the 22nd Conference of the Computer-Human Interaction Special
  Interest Group of Australia on Computer-Human Interaction - OZCHI '10},
  {Margot Brereton}, {Stephen Viller}, {and} {Ben Kraal} (Eds.). {ACM Press},
  New York, New York, USA, 276--279.
\newblock
\showISBNx{9781450305020}
\showDOI{%
\url{http://dx.doi.org/10.1145/1952222.1952281}}


\bibitem{jun2018full}
{Joohee Jun}, {Myeongul Jung}, {So-Yeon Kim}, {and} {Kwanguk~Kenny Kim}. 2018.
\newblock \showarticletitle{Full-Body Ownership Illusion Can Change Our
  Emotion}. In {\em Proceedings of the 2018 CHI Conference on Human Factors in
  Computing Systems}. ACM, 601.
\newblock


\bibitem{kennedy1993simulator}
{Robert~S Kennedy}, {Norman~E Lane}, {Kevin~S Berbaum}, {and} {Michael~G
  Lilienthal}. 1993.
\newblock \showarticletitle{Simulator sickness questionnaire: An enhanced
  method for quantifying simulator sickness}.
\newblock {\em The international journal of aviation psychology\/} {3}, 3
  (1993), 203--220.
\newblock


\bibitem{kilteni2013drumming}
{Konstantina Kilteni}, {Ilias Bergstrom}, {and} {Mel Slater}. 2013.
\newblock \showarticletitle{Drumming in immersive virtual reality: the body
  shapes the way we play}.
\newblock {\em IEEE Transactions on Visualization \& Computer Graphics\/} 4
  (2013), 597--605.
\newblock


\bibitem{kilteni2012sense}
{Konstantina Kilteni}, {Raphaela Groten}, {and} {Mel Slater}. 2012a.
\newblock \showarticletitle{The sense of embodiment in virtual reality}.
\newblock {\em Presence: Teleoperators and Virtual Environments\/} {21}, 4
  (2012), 373--387.
\newblock


\bibitem{kilteni2012extending}
{Konstantina Kilteni}, {Jean-Marie Normand}, {Maria~V Sanchez-Vives}, {and}
  {Mel Slater}. 2012b.
\newblock \showarticletitle{Extending body space in immersive virtual reality:
  a very long arm illusion}.
\newblock {\em PloS one\/} {7}, 7 (2012), e40867.
\newblock


\bibitem{kors2016breathtaking}
{Martijn~JL Kors}, {Gabriele Ferri}, {Erik~D Van Der~Spek}, {Cas Ketel}, {and}
  {Ben~AM Schouten}. 2016.
\newblock \showarticletitle{A breathtaking journey. On the design of an
  empathy-arousing mixed-reality game}. In {\em Proceedings of the 2016 Annual
  Symposium on Computer-Human Interaction in Play}. ACM, 91--104.
\newblock


\bibitem{Krekhov:2018:GVRA}
{Andrey Krekhov}, {Sebastian Cmentowski}, {Katharina Emmerich}, {Maic Masuch},
  {and} {Jens Kr\"{u}ger}. 2018b.
\newblock \showarticletitle{GulliVR: A Walking-Oriented Technique for
  Navigation in Virtual Reality Games Based on Virtual Body Resizing}. In {\em
  Proceedings of the 2018 Annual Symposium on Computer-Human Interaction in
  Play} {\em (CHI PLAY '18)}. ACM, New York, NY, USA, 243--256.
\newblock
\showISBNx{978-1-4503-5624-4}
\showDOI{%
\url{http://dx.doi.org/10.1145/3242671.3242704}}


\bibitem{krekhov2018anim}
{Andrey Krekhov}, {Sebastian Cmentowski}, {and} {Jens Kr{\"u}ger}. 2018a.
\newblock \showarticletitle{VR Animals: Surreal Body Ownership in Virtual
  Reality Games}. In {\em Extended Abstracts Publication of the Annual
  Symposium on Computer-Human Interaction in Play}. ACM, to appear.
\newblock


\bibitem{laviola2000discussion}
{Joseph~J LaViola~Jr}. 2000.
\newblock \showarticletitle{A discussion of cybersickness in virtual
  environments}.
\newblock {\em ACM SIGCHI Bulletin\/} {32}, 1 (2000), 47--56.
\newblock


\bibitem{leite2012shape}
{Lu{\'\i}s Leite} {and} {Veronica Orvalho}. 2012.
\newblock \showarticletitle{Shape your body: control a virtual silhouette using
  body motion}. In {\em CHI'12 Extended Abstracts on Human Factors in Computing
  Systems}. ACM, 1913--1918.
\newblock


\bibitem{lenggenhager2007video}
{Bigna Lenggenhager}, {Tej Tadi}, {Thomas Metzinger}, {and} {Olaf Blanke}.
  2007.
\newblock \showarticletitle{Video ergo sum: manipulating bodily
  self-consciousness}.
\newblock {\em Science\/} {317}, 5841 (2007), 1096--1099.
\newblock


\bibitem{lin2016need}
{Lorraine Lin} {and} {Sophie J{\"o}rg}. 2016.
\newblock \showarticletitle{Need a hand?: how appearance affects the virtual
  hand illusion}. In {\em Proceedings of the ACM Symposium on Applied
  Perception}. ACM, 69--76.
\newblock


\bibitem{BlackWhite}
{{Lionhead Studios}}. 2001.
\newblock \emph{Black \& White}.
\newblock Game [Windows].   (30 March 2001).
\newblock
\newblock
\shownote{Electronic Arts, Redwood City, California, U.S. Last played August
  2003.}


\bibitem{lloyd2007spatial}
{Donna~M Lloyd}. 2007.
\newblock \showarticletitle{Spatial limits on referred touch to an alien limb
  may reflect boundaries of visuo-tactile peripersonal space surrounding the
  hand}.
\newblock {\em Brain and cognition\/} {64}, 1 (2007), 104--109.
\newblock


\bibitem{lombard1997heart}
{Matthew Lombard} {and} {Theresa Ditton}. 1997.
\newblock \showarticletitle{At the heart of it all: The concept of presence}.
\newblock {\em Journal of Computer-Mediated Communication\/} {3}, 2 (1997),
  0--0.
\newblock


\bibitem{lugrin2015anthropomorphism}
{J-L Lugrin}, {Johanna Latt}, {and} {Marc~Erich Latoschik}. 2015.
\newblock \showarticletitle{Anthropomorphism and illusion of virtual body
  ownership}. In {\em Proceedings of the 25th International Conference on
  Artificial Reality and Telexistence and 20th Eurographics Symposium on
  Virtual Environments}. Eurographics Association, 1--8.
\newblock


\bibitem{lugrin2016avatar}
{Jean-Luc Lugrin}, {Ivan Polyschev}, {Daniel Roth}, {and} {Marc~Erich
  Latoschik}. 2016.
\newblock \showarticletitle{Avatar anthropomorphism and acrophobia}. In {\em
  Proceedings of the 22nd ACM Conference on Virtual Reality Software and
  Technology}. ACM, 315--316.
\newblock


\bibitem{maselli2013building}
{Antonella Maselli} {and} {Mel Slater}. 2013.
\newblock \showarticletitle{The building blocks of the full body ownership
  illusion}.
\newblock {\em Frontiers in human neuroscience\/}  {7} (2013), 83.
\newblock


\bibitem{muller2017through}
{Daphne~A Muller}, {Caro~R Van~Kessel}, {and} {Sam Janssen}. 2017.
\newblock \showarticletitle{Through Pink and Blue glasses: Designing a
  dispositional empathy game using gender stereotypes and Virtual Reality}. In
  {\em Extended Abstracts Publication of the Annual Symposium on Computer-Human
  Interaction in Play}. ACM, 599--605.
\newblock


\bibitem{nagel1974like}
{Thomas Nagel}. 1974.
\newblock \showarticletitle{What is it like to be a bat?}
\newblock {\em The philosophical review\/} {83}, 4 (1974), 435--450.
\newblock


\bibitem{normand2011multisensory}
{Jean-Marie Normand}, {Elias Giannopoulos}, {Bernhard Spanlang}, {and} {Mel
  Slater}. 2011.
\newblock \showarticletitle{Multisensory stimulation can induce an illusion of
  larger belly size in immersive virtual reality}.
\newblock {\em PloS one\/} {6}, 1 (2011), e16128.
\newblock


\bibitem{peck2013putting}
{Tabitha~C Peck}, {Sofia Seinfeld}, {Salvatore~M Aglioti}, {and} {Mel Slater}.
  2013.
\newblock \showarticletitle{Putting yourself in the skin of a black avatar
  reduces implicit racial bias}.
\newblock {\em Consciousness and cognition\/} {22}, 3 (2013), 779--787.
\newblock


\bibitem{perez2012my}
{Daniel Perez-Marcos}, {Maria~V Sanchez-Vives}, {and} {Mel Slater}. 2012.
\newblock \showarticletitle{Is my hand connected to my body? The impact of body
  continuity and arm alignment on the virtual hand illusion}.
\newblock {\em Cognitive neurodynamics\/} {6}, 4 (2012), 295--305.
\newblock


\bibitem{petkova2008if}
{Valeria~I Petkova} {and} {H~Henrik Ehrsson}. 2008.
\newblock \showarticletitle{If I were you: perceptual illusion of body
  swapping}.
\newblock {\em PloS one\/} {3}, 12 (2008), e3832.
\newblock


\bibitem{Deadly}
{{Rainbow Studios}}. 2009.
\newblock \emph{Deadly Creatures}.
\newblock Game [Nintendo Wii].   (13 February 2009).
\newblock
\newblock
\shownote{THQ, Agoura Hills, California, U.S. Last played January 2010.}


\bibitem{rhodin2014interactive}
{Helge Rhodin}, {James Tompkin}, {Kwang In~Kim}, {Kiran Varanasi}, {Hans-Peter
  Seidel}, {and} {Christian Theobalt}. 2014.
\newblock \showarticletitle{Interactive motion mapping for real-time character
  control}. In {\em Computer Graphics Forum}, Vol.~33. Wiley Online Library,
  273--282.
\newblock


\bibitem{rhodin2015generalizing}
{Helge Rhodin}, {James Tompkin}, {Kwang~In Kim}, {Edilson De~Aguiar},
  {Hanspeter Pfister}, {Hans-Peter Seidel}, {and} {Christian Theobalt}. 2015.
\newblock \showarticletitle{Generalizing wave gestures from sparse examples for
  real-time character control}.
\newblock {\em ACM Transactions on Graphics (TOG)\/} {34}, 6 (2015), 181.
\newblock


\bibitem{Rigby.2007}
{C.~Scott Rigby} {and} {Richard~M. Ryan}. 2007.
\newblock The Player Experience of Need Satisfaction (PENS): An applied model
  and methodology for understanding key components of the player experience.
\newblock   (2007).
\newblock


\bibitem{riva2014interacting}
{Giuseppe Riva}, {John Waterworth}, {and} {Dianne Murray}. 2014.
\newblock {\em Interacting with Presence: HCI and the Sense of Presence in
  Computer-mediated Environments}.
\newblock Walter de Gruyter GmbH \& Co KG.
\newblock


\bibitem{roth2017alpha}
{Daniel Roth}, {Jean-Luc Lugrin}, {Marc~Erich Latoschik}, {and} {Stephan
  Huber}. 2017.
\newblock \showarticletitle{Alpha IVBO-construction of a scale to measure the
  illusion of virtual body ownership}. In {\em Proceedings of the 2017 CHI
  Conference Extended Abstracts on Human Factors in Computing Systems}. ACM,
  2875--2883.
\newblock


\bibitem{ruddle2009benefits}
{Roy~A Ruddle} {and} {Simon Lessels}. 2009.
\newblock \showarticletitle{The benefits of using a walking interface to
  navigate virtual environments}.
\newblock {\em ACM Transactions on Computer-Human Interaction (TOCHI)\/} {16},
  1 (2009), 5.
\newblock


\bibitem{ryan2000self}
{Richard~M Ryan} {and} {Edward~L Deci}. 2000.
\newblock \showarticletitle{Self-determination theory and the facilitation of
  intrinsic motivation, social development, and well-being.}
\newblock {\em American psychologist\/} {55}, 1 (2000), 68.
\newblock


\bibitem{Ryan.2006}
{Richard~M. Ryan}, {C.~Scott Rigby}, {and} {Andrew Przybylski}. 2006.
\newblock \showarticletitle{The Motivational Pull of Video Games: A
  Self-Determination Theory Approach}.
\newblock {\em Motivation and Emotion\/} {30}, 4 (2006), 344--360.
\newblock
\showISSN{0146-7239}


\bibitem{sanchez2010virtual}
{Maria~V Sanchez-Vives}, {Bernhard Spanlang}, {Antonio Frisoli}, {Massimo
  Bergamasco}, {and} {Mel Slater}. 2010.
\newblock \showarticletitle{Virtual hand illusion induced by visuomotor
  correlations}.
\newblock {\em PloS one\/} {5}, 4 (2010), e10381.
\newblock


\bibitem{Schubert.2018}
{Thomas Schubert}, {Holger Regenbrecht}, {and} {Frank Friedmann}. 2018.
\newblock Igroup Presence Questionnaire (IPQ).
\newblock   (2018).
\newblock
\showURL{%
\url{http://www.igroup.org/pq/ipq/download.php}}


\bibitem{Schubert.2003}
{Thomas~W. Schubert}. 2003.
\newblock \showarticletitle{The sense of presence in virtual environments: A
  three-component scale measuring spatial presence, involvement, and realness}.
\newblock {\em Zeitschrift f{\"u}r Medienpsychologie\/} {15}, 2 (2003), 69--71.
\newblock
\showISSN{1617-6383}


\bibitem{sherman2002understanding}
{William~R Sherman} {and} {Alan~B Craig}. 2002.
\newblock {\em Understanding virtual reality: Interface, application, and
  design}.
\newblock Elsevier.
\newblock


\bibitem{sikstrom2014role}
{Erik Sikstr{\"o}m}, {Amalia De~G{\"o}tzen}, {and} {Stefania Serafin}. 2014.
\newblock \showarticletitle{The role of sound in the sensation of ownership of
  a pair of virtual wings in immersive VR}. In {\em Proceedings of the 9th
  Audio Mostly: A Conference on Interaction With Sound}. ACM, 24.
\newblock


\bibitem{slater2003note}
{Mel Slater}. 2003.
\newblock \showarticletitle{A note on presence terminology}.
\newblock {\em Presence connect\/} {3}, 3 (2003), 1--5.
\newblock


\bibitem{slater2008towards}
{Mel Slater}, {Daniel P{\'e}rez~Marcos}, {Henrik Ehrsson}, {and} {Maria~V
  Sanchez-Vives}. 2008.
\newblock \showarticletitle{Towards a digital body: the virtual arm illusion}.
\newblock {\em Frontiers in human neuroscience\/}  {2} (2008), 6.
\newblock


\bibitem{slater2009inducing}
{Mel Slater}, {Daniel P{\'e}rez~Marcos}, {Henrik Ehrsson}, {and} {Maria~V
  Sanchez-Vives}. 2009.
\newblock \showarticletitle{Inducing illusory ownership of a virtual body}.
\newblock {\em Frontiers in neuroscience\/}  {3} (2009), 29.
\newblock


\bibitem{slater2010first}
{Mel Slater}, {Bernhard Spanlang}, {Maria~V Sanchez-Vives}, {and} {Olaf
  Blanke}. 2010.
\newblock \showarticletitle{First person experience of body transfer in virtual
  reality}.
\newblock {\em PloS one\/} {5}, 5 (2010), e10564.
\newblock


\bibitem{slater1995taking}
{Mel Slater}, {Martin Usoh}, {and} {Anthony Steed}. 1995.
\newblock \showarticletitle{Taking steps: the influence of a walking technique
  on presence in virtual reality}.
\newblock {\em ACM Transactions on Computer-Human Interaction (TOCHI)\/} {2}, 3
  (1995), 201--219.
\newblock


\bibitem{steptoe2013human}
{William Steptoe}, {Anthony Steed}, {and} {Mel Slater}. 2013.
\newblock \showarticletitle{Human tails: ownership and control of extended
  humanoid avatars}.
\newblock {\em IEEE Transactions on Visualization \& Computer Graphics\/} 4
  (2013), 583--590.
\newblock


\bibitem{taylor2005empathy}
{Nicola Taylor} {and} {Tania~D Signal}. 2005.
\newblock \showarticletitle{Empathy and attitudes to animals}.
\newblock {\em Anthrozo{\"o}s\/} {18}, 1 (2005), 18--27.
\newblock


\bibitem{unity}
{Unity Technologies}. 2018.
\newblock {Unity}.
\newblock Website.   (2018).
\newblock
\newblock
\shownote{Retrieved December 12, 2018 from \url{https://unity3d.com/}.}


\bibitem{tsakiris2010my}
{Manos Tsakiris}. 2010.
\newblock \showarticletitle{My body in the brain: a neurocognitive model of
  body-ownership}.
\newblock {\em Neuropsychologia\/} {48}, 3 (2010), 703--712.
\newblock


\bibitem{tsakiris2005rubber}
{Manos Tsakiris} {and} {Patrick Haggard}. 2005.
\newblock \showarticletitle{The rubber hand illusion revisited: visuotactile
  integration and self-attribution.}
\newblock {\em Journal of Experimental Psychology: Human Perception and
  Performance\/} {31}, 1 (2005), 80.
\newblock


\bibitem{Eagle}
{{Ubisoft Montreal}}. 2016.
\newblock \emph{Eagle Flight}.
\newblock Game [VR].   (18 October 2016).
\newblock
\newblock
\shownote{Ubisoft, Montreuil, France. Last played January 2019.}


\bibitem{waltemate2018impact}
{Thomas Waltemate}, {Dominik Gall}, {Daniel Roth}, {Mario Botsch}, {and}
  {Marc~Erich Latoschik}. 2018.
\newblock \showarticletitle{The Impact of Avatar Personalization and Immersion
  on Virtual Body Ownership, Presence, and Emotional Response}.
\newblock {\em IEEE transactions on visualization and computer graphics\/}
  {24}, 4 (2018), 1643--1652.
\newblock


\bibitem{wiemker2015escape}
{Markus Wiemker}, {Errol Elumir}, {and} {Adam Clare}. 2015.
\newblock \showarticletitle{Escape room games}.
\newblock {\em Game Based Learning\/}  {55} (2015).
\newblock


\bibitem{Witmer.2005}
{Bob~G. Witmer}, {Christian~J. Jerome}, {and} {Michael~J. Singer}. 2005.
\newblock \showarticletitle{The Factor Structure of the Presence
  Questionnaire}.
\newblock {\em Presence: Teleoperators and Virtual Environments\/} {14}, 3
  (2005), 298--312.
\newblock
\showISSN{1054-7460}
\showDOI{%
\url{http://dx.doi.org/10.1162/105474605323384654}}


\bibitem{witmer1998measuring}
{Bob~G Witmer} {and} {Michael~J Singer}. 1998a.
\newblock \showarticletitle{Measuring presence in virtual environments: A
  presence questionnaire}.
\newblock {\em Presence\/} {7}, 3 (1998), 225--240.
\newblock


\bibitem{Witmer.1998}
{Bob~G. Witmer} {and} {Michael~J. Singer}. 1998b.
\newblock \showarticletitle{Measuring Presence in Virtual Environments: A
  Presence Questionnaire}.
\newblock {\em Presence: Teleoperators and Virtual Environments\/} {7}, 3
  (1998), 225--240.
\newblock
\showISSN{1054-7460}
\showDOI{%
\url{http://dx.doi.org/10.1162/105474698565686}}


\bibitem{won2015homuncular}
{Andrea~Stevenson Won}, {Jeremy~N Bailenson}, {and} {Jaron Lanier}. 2015.
\newblock \showarticletitle{Homuncular flexibility: the human ability to
  inhabit nonhuman avatars}.
\newblock {\em Emerging Trends in the Social and Behavioral Sciences: An
  Interdisciplinary, Searchable, and Linkable Resource\/} (2015), 1--16.
\newblock


\bibitem{8643070}
{W. {Xu}}, {A. {Chatterjee}}, {M. {Zollhofer}}, {H. {Rhodin}}, {P. {Fua}}, {H.
  {Seidel}}, {and} {C. {Theobalt}}. 2019.
\newblock \showarticletitle{$Mo^{2}Cap^{2}$ : Real-time Mobile 3D Motion
  Capture with a Cap-mounted Fisheye Camera}.
\newblock {\em IEEE Transactions on Visualization and Computer Graphics\/}
  (2019), 1--1.
\newblock
\showISSN{1077-2626}
\showDOI{%
\url{http://dx.doi.org/10.1109/TVCG.2019.2898650}}


\bibitem{yee2007proteus}
{Nick Yee} {and} {Jeremy Bailenson}. 2007.
\newblock \showarticletitle{The Proteus effect: The effect of transformed
  self-representation on behavior}.
\newblock {\em Human communication research\/} {33}, 3 (2007), 271--290.
\newblock


\end{thebibliography}
